\documentclass[%
reprint,
superscriptaddress,
amsmath,amssymb,
aps,prl
]{revtex4-1}
\usepackage[utf8]{inputenc}
\usepackage[utf8]{inputenc}
\usepackage[english]{babel}
\usepackage{amsmath}
\usepackage{siunitx}
\usepackage{graphicx}
\usepackage{color}
\usepackage{ulem}

\begin{document}
\title{Fluorescence enhancement by dark plasmon modes}

\author{Manuel Peter}
\affiliation{Physikalisches Institut, Nussallee 12, Universität Bonn, 53115 Bonn, Germany}

\author{Julia F.~M.~Werra}
\affiliation{Humboldt-Universität zu Berlin, Institut für Physik, AG Theoretische Optik \& Photonik, Newtonstr. 15, 12489 Berlin, Germany}

\author{Cody Friesen}
\affiliation{Physikalisches Institut, Nussallee 12, Universität Bonn, 53115 Bonn, Germany}

\author{Doreen Achnitz}
\affiliation{Physikalisches Institut, Nussallee 12, Universität Bonn, 53115 Bonn, Germany}

\author{Kurt Busch}
\affiliation{Humboldt-Universität zu Berlin, Institut für Physik, AG Theoretische Optik \& Photonik, Newtonstr. 15, 12489 Berlin, Germany}
\affiliation{Max-Born-Institut, Max-Born-Str. 2A, 12489 Berlin, Germany}

\author{Stefan Linden}
\email{linden@physik.uni-bonn.de}
\affiliation{Physikalisches Institut, Nussallee 12, Universität Bonn, 53115 Bonn, Germany}


\begin{abstract}
We investigate the fluorescence properties of colloidal quantum dots attached to gold rod nanoantennas. 
These structures are fabricated by a two step electron beam lithography process in combination with a chemical linking method. 
By varying the nanoantenna length, the plasmon modes of the nanoantennas are successively tuned through the emission band of the quantum dots.
We observe a pronounced fluorescence enhancement both for short and long nanoantennas. These findings can be attributed to the coupling of the quantum dots to bright and dark plasmon modes, respectively. 
\end{abstract}

\maketitle

The optical properties of noble metal nanostructures in the visible and the near infrared spectral region are dominated by collective excitations of the conduction band electrons~\cite{Maier_Plasmonics}. 
These so-called plasmon modes can be classified according to their radiative character into two groups. Bright plasmon modes are of dipolar character and couple efficiently to far field radiation.
For instance, the fundamental plasmon modes of metallic nanospheres~\cite{Maier_Plasmonics}, nanorods~\cite{Novotny2007}, split-ring resonators~\cite{Rockstuhl2006}, and bow-tie antennas~\cite{kinkhabwala2009} belong to this group.
In contrast, dark plasmon modes exhibit a vanishing dipole moment and, as a result, their radiative damping is strongly suppressed.
Examples include higher-order multipole  resonances supported by individual metal  nanoparticles and antisymmetric modes in coupled plasmonic nanostructures ~\cite{yang2010plasmon,Gomez2013}.

Since the pioneering experiments of K.~H.~Drexhage in the late 1960s, it has been known that the emission properties of quantum emitters can be modified by a nearby metal interface~\cite{Drexhage1968}. Corresponding experiments with quantum emitters in the vicinity of metal  nanostructures have shown that the electromagnetic coupling of an emitter to plasmon modes of the metal nanostructure can lead both to significant enhancement and to quenching of the fluorescence intensity~\cite{anger2006,kuhn2006,muskens2007,kinkhabwala2009}. In this regard, the outcome depends critically on the spatial arrangement and the separation of the emitter and the metal nanostructure as well as the radiative character of the involved plasmon modes. 

Several different effects can lead to an increase of the collected fluorescence intensity~\cite{koenderink2017}: (i) The enhancement of the pump intensity at the position of the quantum emitter due to a strong near-field associated with the excitation of a bright plasmon mode by the pump beam can increase the excitation rate. (ii) Metal nanostructures can serve as antennas such that the radiation efficiency of the combined system is larger than that of the bare quantum emitter (Purcell effect). (iii) Moreover, the spatial radiation apttern of the emitter can be modified by the metal nanostructure so that a larger fraction of the emitted light enters the collection optics.
Quenching of the fluorescence is usually attributed to the coupling of the quantum emitter to dark plasmon modes. The nonradiative decay of a dark plasmon mode results in the conversion of the excitation energy into heat. Quenching is usually the dominant process for very small separations (typically below 5-10 nm) between the quantum emitter and the metal nanostructure.  

In this letter, we investigate the coupling of colloidal semiconductor quantum dots to gold rod nanoantennas by fluorescence microscopy. The rod antennas feature a series of bright and dark plasmon modes~\cite{Schider2003,taminiau2011optical}. By varying the length of the nanoantennas we can tune these modes one by one through the emission band of the quantum dots. A strong enhancement of the fluorescence intensity is observed if the fundamental plasmon mode overlaps with the emission spectrum of the dots. 
Interestingly, enhanced fluorescence intensity is also observed for longer antennas whose scattering spectra feature no bright plasmonic mode resonant to the quantum dot emission. We attribute this finding to the coupling of  the quantum dots to the dark second order plasmon mode of these nanoantennas. For symmetry reasons, this mode with an antisymmetric current distribution can not be observed in the scattering spectra recorded under normal incidence. However, it can be excited by the quantum dots in the near-field and radiate to the far-field with a quadrupole radiation pattern \cite{taminiau2011optical,Kajetan2012}. This interpretation is supported by numerical calculations of the radiation rate of the coupled system.

\begin{figure}[htbp]
	\centering
	\fbox{\includegraphics[width=\linewidth]{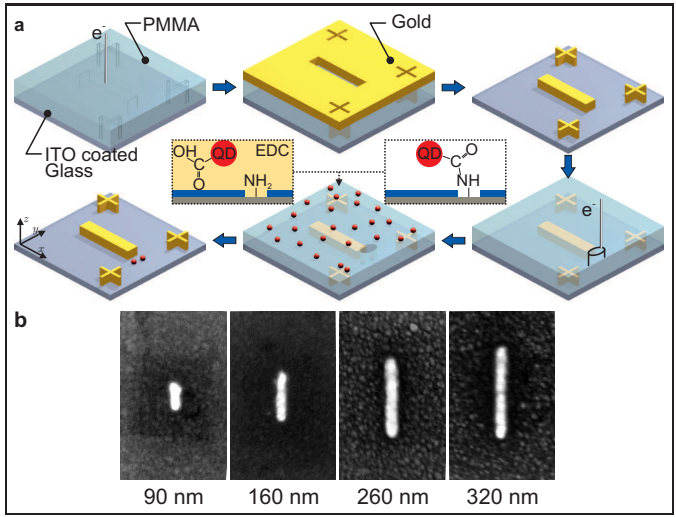}}
	\caption{(a) Scheme of the sample fabrication process. (b) Scanning electron micrographs of four metal nanoantennas with $L=\SI{90}{\nano \metre}$, $L=\SI{160}{\nano \metre}$, $L=\SI{260}{\nano \metre}$, and $L=\SI{320}{\nano \metre}$, respectively.}
	\label{Fig_1}
\end{figure}

\begin{figure}[tttt]
	\centering
	\fbox{\includegraphics[width=\linewidth]{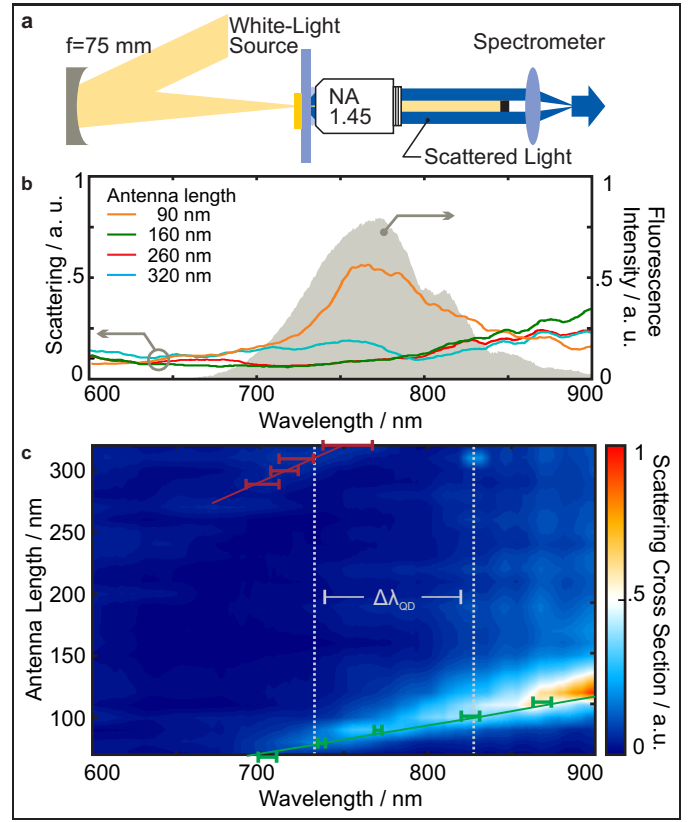}}
	\caption{(a) Scheme of the dark field spectroscopy setup. (b) Measured scattering spectra of the four nanoantennas depicted in Fig.~\ref{Fig_1}. Additionally, the fluorescence spectrum of the quantum dots is depicted in grey. (c) Measured scattering spectra of the complete set of nanoantennas presented on a false color scale. The quantum dot emission band is indicated by the dotted vertical lines.}
	\label{Fig_2}
\end{figure}

The fabrication of the samples follows the process shown schematically in Fig.~\ref{Fig_1}(a). We first produce a set of gold rod nanoantennas on a microscope cover slip coated with $\SI{8}{\nano \metre}$ of indium tin oxide (ITO) by standard electron-beam lithography (EBL) with polymethyl methacrylate (PMMA) as positive tone electron resist and thermal evaporation of gold. 
Within the set, the nanoantenna length $L$ is increased from $L=\SI{70}{\nano \metre}$ to $L=\SI{320}{\nano \metre}$ in steps of $\SI{10}{\nano \metre}$.
The height $h$ and the width $w$ of the gold nanoantennas are chosen to be $h=\SI{40}{\nano \metre}$ and $w=\SI{30}{\nano \metre}$, respectively. 
Fig.~\ref{Fig_1}(b) depicts scanning electron micrographs of four selected antennas. 
We use commercial CdSeTe quantum dots  that are shelled with ZnS and coated with a polymer providing carboxyl surface groups (Qdot 800 Carboxyl Quantum Dots, Thermo Fisher Scientific). 
The emission spectrum of an ensemble of bare quantum dots is centered at $\SI{780}{\nano \metre}$ wavelength and has a full-width at half maximum of approximately $\SI{100}{\nano \metre}$ (see Fig.~\ref{Fig_2} (b)).
Its fluorescence shows no preferential polarization direction since the quantum dot orientations are randomly distributed in the ensemble. 
To deposit quantum dots at the antennas' tips, we use a second lithographic step in combination with chemical linking \cite{Peter2017}. For this purpose, the sample with the gold antennas is coated again with PMMA and for each nanoantenna a $\SI{70}{\nano \metre} \times \SI{70}{\nano \metre}$ large rectangular area centered at the tip is defined by exposure with the electron beam.
After development, the PMMA film with the holes serves as a template for the subsequent chemical surface functionalization. To this end, the sample is placed for an hour in a solution of ${10~\%}$ (3-Aminopropyl)triethoxysilane (APTES) in isopropyl alcohol to silanize the ITO layer in the unveiled patches. Next, 1-Ethyl-3-(3-dimethylaminopropyl) carbodiimide (EDC) is added to the aqueous pH buffered quantum dot solution and the substrate is immersed for two hours with constant stirring in this solution. EDC acts as an activating agent that mediates the chemical link between the carboxyl surface groups of the quantum dots and the silanized substrate. After rinsing the substrate with deionised water, the PMMA mask is finally removed in a lift-off process and the quantum dots stick to the modified surface areas. 
The quantum dot patches defined in this way contain approximately 10 quantum dots. 

Scattering spectra of the individual nanoantennas are measured by normal incidence dark-field spectroscopy \cite{Weigel2014}.
A supercontinuum white-light laser is focused on the sample with a concave mirror (see fig.~\ref{Fig_2}(a)). The linear polarization of the beam is chosen to be parallel to the long axis of the nanoantennas. The transmitted white-light beam as well as scattered light from the gold nanoantennas are collected with an oil immersion microscope objective (numerical aperture NA=1.45). 
A small circular stop blocks the white light behind the objective.
Since the angular distribution of the scattered light is considerably broader than the divergence of the white light beam, a large fraction of the scattered light passes this stop. 
An adjustable knife-edge aperture placed in an intermediate image plane is used to select single nanoantennas.
The corresponding scattering spectra are recorded with a CCD camera attached to a grating spectrometer and referenced to the incident white-light spectrum.

Figure~\ref{Fig_2} (b) shows the measured scattering scattering spectra of the four nanoantennas depicted above. The fluorescence spectrum of the quantum dots (grey area) is included for latter reference.
Additionally, scattering spectra of the complete set of nanoantennas are presented on a false color scale in Fig.~\ref{Fig_2} (c).
All spectra are normalized to the to the overall maximum of the set.
The antenna with length $L=\SI{90}{\nano \metre}$ exhibits a pronounced resonance at a wavelength of $\SI{760}{\nano \metre}$. This resonance corresponds to the fundamental antenna mode, which is a bright plasmon with dipolar character.   
With increasing antenna length $L$, the fundamental antenna  mode shifts to larger wavelength (see green symbols in Fig.~\ref{Fig_2} (c)) and the scattering strength increases \cite{Novotny2007}. 
For antennas larger than $\SI{140}{\nano \metre}$, the fundamental resonance wavelength lies outside our measurement range. 
An additional, faint resonance at around $\SI{760}{\nano \metre}$ wavelength can be observed for the $\SI{320}{\nano \metre}$ long nanoantenna. 
We attribute this resonance to the third order mode of the antenna, which is a bright plasmon mode. 
Like the fundamental mode, the resonance peak of the third order mode shifts to longer wavelength with increasing antenna length (see red symbols in Fig.~\ref{Fig_2} (c)).

\begin{figure}[tttt]
	\centering
	\fbox{\includegraphics[width=\linewidth]{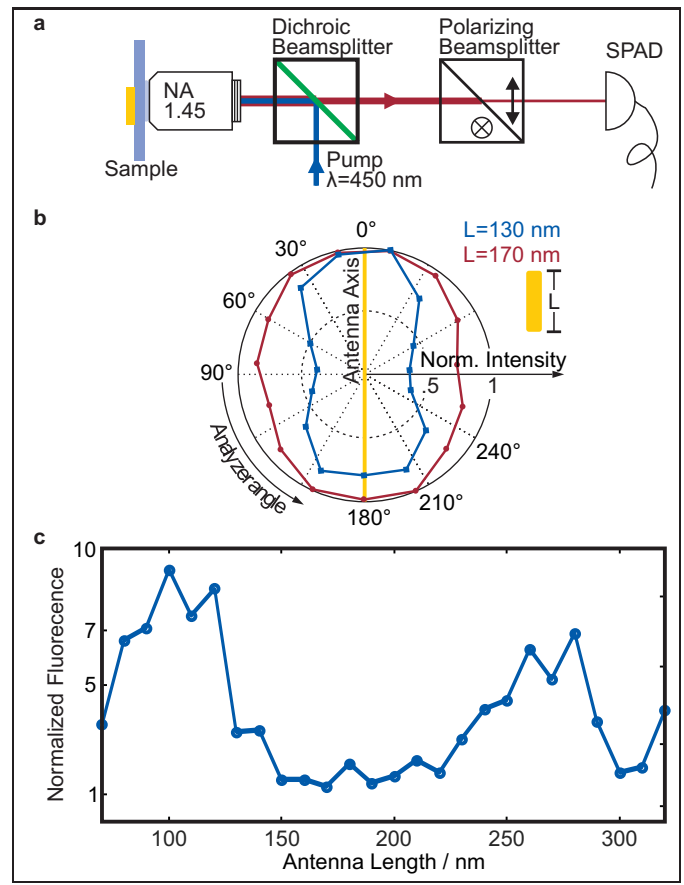}}
	\caption{(a) Scheme of the confocal fluorescence microscopy setup.(b) Polarization plots of the fluorescence for two different antenna lengths (blue symbols: $L=\SI{130}{\nano \metre}$, red symbols: $L=\SI{170}{\nano \metre}$). The curves serve as guides to the eye. (c) Fluorescence intensity of the quantum dots as a function of the antenna length.}
	\label{Fig_3}
\end{figure}

The emission properties of the quantum dots coupled to the nanoantennas are investigated by confocal fluorescence microscopy (see Fig.~\ref{Fig_3} (a)).
For this purpose, a blue pump laser ($\lambda=\SI{450}{\nano \metre}$) is focused by a high-numerical-aperture objective ($100 \times$ magnification, NA=1.45) through the substrate onto a single antenna to excite the quantum dots at the tip. The emitted fluorescence is collected with the same objective and separated from reflected pump light by dichroic filters and detected either with a spectrometer or a single photon avalanche diode. A polarizer in front of the detector is used to analyze the polarization properties of the fluorescence.

Figure ~\ref{Fig_3} (b) shows polarization plots of the quantum dot fluorescence for two different antenna lengths ($L=\SI{130}{\nano \metre}$ and $L=\SI{170}{\nano \metre}$). 
In this figure, the fluorescence signal for each antenna is normalized to its respective maximum value.
The fundamental plasmon resonance  of the $\SI{130}{\nano \metre}$ long nanoantenna overlaps with the emission band of the quantum dots ranging from $\SI{730}{\nano \metre}$ to $\SI{830}{\nano \metre}$. The fluorescence of the quantum dots attached to this antenna features a clear preferential polarization along the antenna axis (see blue symbols in Fig.~\ref{Fig_3} (b)). 
This indicates that the quantum dot emission is mediated by the nanoantenna ~\cite{Ming2011}.
The second antenna is $\SI{170}{\nano \metre}$ long and its fundamental resonance is red-shifted with respect to the quantum dot emission band. In contrast to the previous case, the fluorescence is nearly unpolarized for this non-resonant nanoantenna (see red symbols in Fig.~\ref{Fig_3} (b)). Obviously, the non resonant antenna modifies the quantum dot emission to a significantly lesser extend than the resonant one. 

Figure~\ref{Fig_3} (c) displays the fluorescence intensity of the quantum dots as a function of the antenna length.
Each data point corresponds to the average of three nominally identical realizations and is normalized to the fluorescence of an equivalent patch of bare quantum dots (without nanoantenna). For all measurements, the polarizer is  oriented along the antenna axis.
We verified that the fluorescence intensity of the bare gold nanoantennas  without quantum dots is negligible for the chosen excitation laser power (not shown).
Enhanced fluorescence (compared to a patch of quantum dots without antenna) is observed for two sets of antennas: (i) short antennas with lengths ranging from $L=\SI{70}{\nano \metre}$ to $L=\SI{140}{\nano \metre}$ and (ii) long antennas with $L$ between $L=\SI{240}{\nano \metre}$ and $L=\SI{290}{\nano \metre}$.
The fluorescence enhancement in the former case is in line with our expectations, since the short nanoantennas are resonant with the quantum dot emission band and the corresponding first antenna mode is a bright plasmon. 
In contrast, the fluorescence enhancement observed for the long antennas is at first sight surprising. 
The scattering spectra of these antennas show no resonance in the quantum dot emission band.
Their fundamental mode is considerably red shifted with respect to the quantum dots and the next bright plasmon mode, the third order antenna mode, only overlaps for even longer antennas ($L\ge\SI{290}{\nano \metre}$) with the emission band.
Hence, we exclude these two bright plasmon modes as the cause of the fluorescence enhancement.
Instead, we attribute the emission enhancement to the coupling of the quantum dots to the dark second order plasmon mode of the antennas. 
This mode features an antisymmetric current distribution and, hence, can not been seen in the far field scattering spectra recorded under normal incidence. However, it can be excited by sources in the near-field~\cite{Kajetan2012}, i.e, by the quantum dots, and subsequently radiate to the far-field featuring a quadrupole radiation pattern. 

To support our interpretation of the experimental data, we performed numerical calculations using the discontinuous Galerkin time-domain method (DGTD)~\cite{Busch2011}. 
Within the DGTD method, the system is discretized into a tetrahedral mesh where, on each element of the mesh, the electromagnetic field is expanded into fourth-order basis functions. The glass substrate is modeled as a dispersion- and lossless dielectric with refractive index $n=1.53$ while gold is described via a Drude-Lorentz model whose parameters are fitted directly to experimental data \cite{Johnson_1972}.
From this setup, all relevant quantities such as field distributions, scattering spectra, densities of states, and emission rates can be computed in a straightforward manner (see, e.g., Ref.~\cite{Schell2014}).

\begin{figure}[ttt]
	\centering
	\fbox{\includegraphics[width=\linewidth]{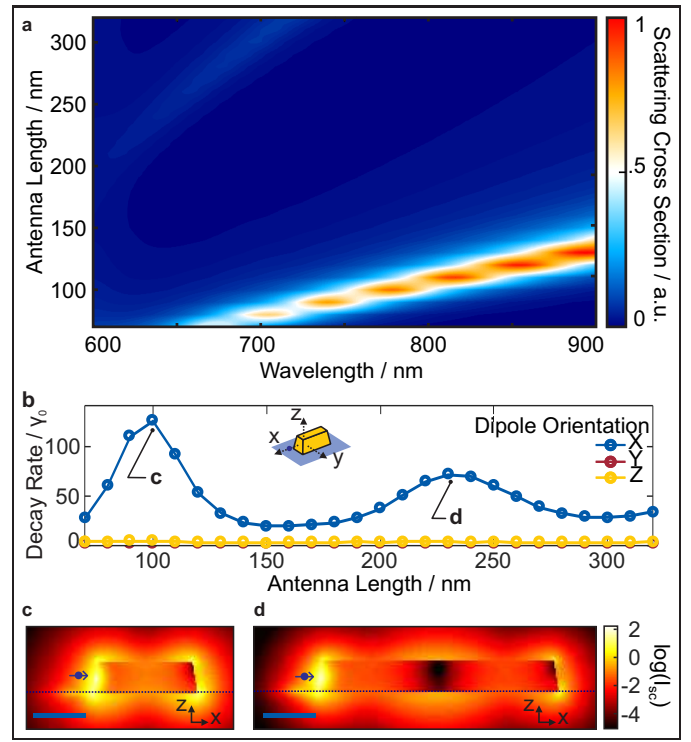}}
	\caption{(a) Calculated scattering spectra of gold nanoantennas as a function of antenna length presented on a false color scale. (b) Calculated decay rate enhancement of a dipole positioned $\SI{10}{\nano \metre}$ in front of a gold nanoantenna (see dot in inset) as a function of antenna length. The three curves correspond to three different dipole orientations along the coordinate axis (data of y orientation hidden behind z orientation data). (c) Calculated near-field distribution of the $\SI{100}{\nano \metre}$ long nanoantenna excited by the dipole. (d) Calculated near-field distribution of the $\SI{230}{\nano \metre}$ long nanoantenna excited by the dipole. The blue scale bars are $\SI{50}{\nano \metre}$ long. The position of the dipole and its orientation are indicated by the dot and the arrow, respectively.}
	\label{Fig_4_einspaltig}
\end{figure}

The calculated scattering spectra shown in Fig.~\ref{Fig_4_einspaltig} (a) qualitatively reproduce all experimental trends. 
We observe a pronounced resonance for the short nanoantennas ($L\le\SI{140}{\nano \metre}$) and an additional fainter resonance for long ($L\ge\SI{200}{\nano \metre}$) nanoantennas. The latter overlaps with the quantum dot emission band only for $L\ge\SI{300}{\nano \metre}$.
Our assignment of these resonances to the first and third order plasmon mode, respectively, is confirmed by the calculated near-field distributions of these modes (not shown).
Like in the experimental data, the dark second order mode of the nanoantennas does not emerge in the calculated scattering spectra. 

Figure~\ref{Fig_4_einspaltig} (b) displays the calculated radiative decay rate enhancements of a dipole with the experimentally measured quantum dot fluorescence spectrum (see Fig. ~\ref{Fig_2} (b)) positioned $\SI{10}{\nano \metre}$ in front of the tip of a gold nanoantenna compared to a dipole in vacuum for three dipole orientations. 
A strong influence of the nanoantenna can be only seen for the dipole oriented along the nanoantenna axis ($x$-axis). Here, we find a significant 
radiative decay rate enhancement for two sets of antennas: (i) short antennas with lengths ranging from $L=\SI{80}{\nano \metre}$ to $L=\SI{120}{\nano \metre}$ and (ii) long antennas with lengths between $L=\SI{220}{\nano \metre}$ and $L=\SI{260}{\nano \metre}$. To identify the involved plasmonic modes, we calculated the near-field distribution $\log_{10}(I_{\rm sc})$ ($I_{\rm sc}=|\mathbf{E}-\mathbf{E}_{\textrm{vac}}|^2$, where $\mathbf{E}$ is the electric field of the dipole with antenna and substrate and $\mathbf{E}_{\rm vac}$ is the electric field of the dipole in vacuum) for two representative antennas with $L=\SI{100}{\nano \metre}$ and a $L=\SI{230}{\nano \metre}$, respectively. The nanoantennas are excited by a dipole oriented along the $x$-axis with an emission wavelength of $\SI{780}{\nano \metre}$ positioned $\SI{10}{\nano \metre}$ in front of the tip of the antenna. The field distribution of the short nanoantenna (see Fig.~\ref{Fig_4_einspaltig} (c)) has no nodal plane inside of the nanoantenna and hence corresponds to the bright fundamental plasmon mode.
In contrast, the field distribution of the long nanoantenna (see Fig.~\ref{Fig_4_einspaltig} (d)) features  one nodal plane inside of the nanoanatenna, which is a clear finger print of the dark second order plasmon mode.
These findings support our interpretation of the experimental  data that the fluorescence enhancement of the short nanoantennas can be attributed to the bright fundamental plasmon mode while that observed for the long antennas can be traced back to the coupling of the quantum dots to the dark second order plasmon mode. 

In summary, we investigate the fluorescence properties of colloidal quantum dots attached to gold rod nanoantennas. Our experiments and our numerical calculations show that enhancement of the fluorescence intensity can result from a coupling of the quantum dots to both the fundamental and the second order plasmon mode. The latter observation is surprising since the second order plasmon mode is usually considered as a dark mode and hence is expected to lead to quenching of the fluorescence.

The authors declare no competing financial interest.

	S.L. acknowledgse financial support through DFG TRR 185 and by the German Federal Ministry of Education and Research through the funding program Photonics Research Germany (project 13N14150). K.B. acknowledges financial support by the German Federal Ministry of Education and Research through the funding program Photonics Research Germany (project 13N14149).

\bibliography{plasmonic}

\end{document}